\documentclass[aps,pra,reprint,showpacs,superscriptaddress, nofootinbib,onecolumn ]{revtex4-2}

\usepackage{graphicx}
\usepackage{dcolumn}
\usepackage{bm}
\usepackage{xcolor}  
\expandafter\let\csname equation*\endcsname\relax
\expandafter\let\csname endequation*\endcsname\relax
\usepackage{amsmath}
\usepackage{graphicx}
\usepackage{soul}
\usepackage{float}
\usepackage{physics}
\emergencystretch=\maxdimen
\usepackage[export]{adjustbox}

\begin{document}

\title{From Feynman-Vernon to Wiener Stochastic Path Integral}

\author{Antonio \surname{Camurati}}
\affiliation{Departamento de F\'isica, PUC-Rio, 22452-970, Rio de Janeiro RJ, Brazil}

\author{Felipe \surname{Sobrero}}
 \affiliation{Centro Brasileiro de Pesquisas Físicas, 22290-180, Rio de Janeiro RJ, Brazil}
\author{Bruno \surname{Suassuna}}
\affiliation{Departamento de Matem\'atica, PUC-Rio, 22452-970, Rio de Janeiro RJ, Brazil}
 \author{Pedro \surname{V.~Paraguass\'u}}
 \affiliation{Departamento de F\'isica, PUC-Rio, 22452-970, Rio de Janeiro RJ, Brazil}


\begin{abstract}
We establish a direct connection between the Feynman-Vernon path integral formalism for open quantum systems and the Wiener path integral used in classical stochastic dynamics. By considering a generalized influence functional in the strong decoherence limit, we demonstrate that integrating over the quantum coherence length leads to a derivation of stochastic Langevin dynamics. Specifically, we show that the quantum Feynman measure transforms into the stochastic Wiener measure. Applying this framework to the Wigner function representation, we show that the system follows a stochastic path interpretable via classical probability theory. Finally, we address the inverse problem: constructing an equivalent quantum influence functional from a given classical Langevin equation.
 \end{abstract}

\maketitle

\section{Introduction}
The Feynman-Vernon formalism remains a cornerstone in the description of open quantum systems, allowing one to trace out environmental degrees of freedom via path integrals. Recently, this technique has seen a resurgence, finding applications ranging from low-energy quantum gravity \cite{parikh2021signatures, parikh2021quantum, chawla2023quantum, parikh2025quantum, cho2023graviton} and optomechanics \cite{paraguassu2024quantum, paraguassu2025apparent, sobrero2025response}  to quantum thermodynamics \cite{funo2017path, funo2018path, qiu2020path}. Parallel to this, path integral methods have become indispensable in classical statistical physics. From stochastic thermodynamics \cite{paraguassu2023heat, paraguassu2023heat2, chernyak2006path, melo2024brownian, paraguassu2024brownian, chatterjee2010exact, taniguchi2008inertial} to rigorous mathematical formulations \cite{arenas2010functional, barci2016path, moreno2019conditional,cugliandolo2019building, marguet2021supersymmetries}, these methods provide a powerful framework for investigating stochastic phenomena.

A central theme at the intersection of these fields is the emergence of classical stochastic dynamics from underlying quantum subsystems. While approaches such as hybrid master equations \cite{layton2024healthier, oppenheim2023objective} and post-quantum theories \cite{oppenheim2023postquantum, sajjad2024quantum, oppenheim2023path, layton2024classical} have been explored, a distinct dynamics is the Quantum Induced Stochastic Dynamics (QISD) \cite{parikh2021signatures, paraguassu2024quantum, calzetta1994noise}. This approach typically relies on a saddle-point approximation of the Feynman-Vernon effective action, assuming a equivalence of forward and backward trajectories \cite{parikh2021signatures, paraguassu2024quantum, weiss2012quantum, feynman2000theory, kamenev2023field}. This procedure yields a Langevin-like equation describing the stochastic dynamics induced by the quantum interaction \cite{weiss2012quantum, feynman2000theory, caldeira1983path}. 

Furthermore, once a Langevin-like equation is established, the stochastic dynamics can be rigorously treated using the Martin-Siggia-Rosen (MSR) formalism \cite{martin1973statistical, wio2013path}, which employs a Wiener path integral measure \cite{chaichian2018path,wio2013path}. This raises a fundamental question regarding the rigorous mathematical bridge between the oscillatory Feynman measure of quantum mechanics and the diffusive Wiener measure of stochastic processes, given their distinct natures. While the link between these two dynamics has been explored in \cite{caldeira1983path, weiss2012quantum, kamenev2023field}, a rigorous definition of the measure—particularly regarding its connection with multiplicative noise—is still lacking.

In this article, we bridge this gap by considering a generalized influence functional that allows for a perturbative treatment of the measures. We demonstrate that, in the strong decoherence limit, the Feynman measure naturally transforms into a Wiener measure upon integrating out the coherence length. This establishes a rigorous connection between open quantum dynamics and stochastic processes at the path integral level.

The paper is structured as follows: Sec.~\ref{sec: feynmantowiener} establishes the connection between Feynman and Wiener measures using the generalized influence functional. In Sec.~\ref{sec: estochastic evol}, we apply this framework to the Wigner function, showing that its evolution is consistent with a classical stochastic process. Section~\ref{sec: quantization} discusses the inverse problem of quantizing Langevin equations, by constructing equivalent influence functionals. Concluding remarks are given in Sec.~\ref{sec: conclusion}

\section{From Feynman measure to Wiener measure}\label{sec: feynmantowiener}

We begin with a review of the Feynman-Vernon formalism. Consider a bipartite composite system consisting of a system of interest $C$ coupled to another system $Q$, the environment. We consider system $C$ as a particle in one-dimension, with coordinates $q$, and the system $Q$ is a set of particles described by coordinates $Q=\{Q_j\}$. The case where the system $C$ is a set of multiple particles is analogous. We note that in applications to quantum optics, optomechanics or in describing interactions of detectors with gravitational waves, what we call a particle in one-dimension could model an optical mode in a cavity or a mode of the gravitational wave field \cite{paraguassu2024quantum, parikh2021signatures, caldeira1983path}.

The total Hamiltonian is given by $H = H_C + H_Q + H_{\text{int}}$, where $H_C$ and $H_Q$ describe the dynamics of each system in the absence of interaction and $H_{\text{int}}$ an interaction term, such as a potential $V_{\mathrm{int}}(q,Q)$.

We are interested in the reduced dynamics of system $C$, which is obtained by tracing out the environment's degrees of freedom from the total density matrix:
\begin{equation}
\rho_{\text{tot}}(t)=\exp(-\frac{i}{\hbar}Ht)\rho_{\mathrm{tot}}(0)\exp(\frac{i}{\hbar}Ht).   
\end{equation}
Assuming that at the initial time $t=0$ the total system is in a separable initial state $\rho_{\text{tot}}(0) = \rho_0(q_0, q_0') \otimes \rho_Q$, the time evolution of the reduced density matrix $\rho(q_t, q_t', t) = \text{Tr}_Q [\rho_{\text{tot}}(t)]$ can be expressed in terms of a path integral over paths $q(t)$ and $q'(t)$ with the appropriate boundary conditions. 

The evolution to time $t=\tau$ is given by
\begin{equation}
    \rho (q_\tau, q_\tau', \tau) = \int dq_0 dq_0' \rho_0(q_0, q_0') \mathcal{J}_\tau[q_\tau,q_\tau'|q_0, q_0'],
    \end{equation}
where $\mathcal{J}_\tau$ is the effective propagator that describes the dynamics of the open system. According to the Feynman-Vernon formalism~\cite{feynman2000theory}, the superpropagator $\mathcal{J}_\tau$ given by
\begin{equation}
    \mathcal{J}_\tau[q_\tau,q_\tau'|q_0, q_0'] = \int \mathcal{D}q \mathcal{D}q' \exp\left(\frac{i}{\hbar}\int_0^\tau dt \left(L_q - L_{q'}\right)\right)\mathcal{F}[q,q'],
    \label{eq:Fundamental FV}
\end{equation}
where $\mathcal{F}=\mathcal{F}_\tau$ is the Feynman-Vernon influence functional.

In the absence of interactions, we have $\mathcal{F}=1$ and the superprogator is the product of propagators $\bra{q_\tau}U_C(\tau)\ket{q_0}\bra{q_\tau'}U_C(\tau)^\dagger\ket{q_0'}$, where $U_C(\tau)=\exp(-\frac{i}{\hbar}H_c\tau)$, and we use the path-integral representations
\begin{equation}
    \bra{q_\tau}U_C(\tau)\ket{q_0} = \int \mathcal{D}q \exp(\frac{i}{\hbar}\int_0^{\tau} dt L_q)\ \ , \  \ \bra{q_\tau'}U(\tau)^\dagger\ket{q_0'} =  \bra{q_0'}U_C(\tau)\ket{q_\tau'}^* = \int \mathcal{D}q' \exp(-\frac{i}{\hbar}\int_0^{\tau} dt L_q'),
\end{equation}
where the formal measures $\mathcal{D}q'$ and $\mathcal{D}q$ are Feynman measures, defined as \cite{chaichian2018path}
\begin{equation}
    \mathcal{D}q = \prod_{t=0}^\tau  \frac{dq(t)}{\sqrt{\frac{2i\pi\hbar}{m}dt}}, \;\; \mathcal{D}q' = \prod_{t=0}^\tau  \frac{dq'(t)}{\left(\sqrt{\frac{2i\pi\hbar}{m}dt}\right)^*}. \label{eq:fey_meas}
\end{equation}
We notice that as $t$ ranges from $[0,\tau]$, the paths $q(t)$ go from $q_0$ to $q_\tau$ and the paths $q'(t)$ go backwards from $q_\tau'$ to $q_0'$.

The notation in \eqref{eq:fey_meas} indicates that the path integrals over $\mathcal{D}q$ are defined as the $N\to+\infty$ limit of integrals over $(N-1)$-dimensional Euclidean space with volume measure $dq_1\cdots dq_{N-1}$ weighted by a constant $A_N=(\frac{2i\pi\hbar\Delta t}{m})^{-(N-1)/2}$, where $\Delta t=\tau/N$ and we have partitioned the interval $[0,\tau]$ into equal parts of size $\Delta t$, replacing the paths $q=q(t)$ by their evaluation $q_k=q(t_k)$ at times $t_k=k\Delta t$ for $k=0,\ldots,N$, keeping the end-points fixed at the values prescribed by the boundary conditions. In this prescription, the integrand of a path-integral, which is a functional of paths $\Phi[q]$, must be replaced by a function $\Phi_N(q_0,q_1,\ldots,q_N)$ of positions at discrete times. The path-integrals over $\mathcal{D}q'$ are defined in the same way, but normalized with the complex conjugate $A_N^*$.

The difference in normalization between the Feynman measures $\mathcal{D}q$ and $\mathcal{D}q'$ follows from the derivation of \eqref{eq:Fundamental FV}, since both propagators $\bra{q,Q}U(t)\ket{q'',Q''}$ and their complex conjugate $\bra{q'',Q''}U(t)^{\dagger}\ket{q',Q'} = \bra{q',Q'}U(t)\ket{q'',Q''}^*$ appear in the expression for the evolution of the total density matrix, from where the partial trace with respect to the environment $Q$ gives \eqref{eq:Fundamental FV}, with formal measures $\mathcal{D}q$ and $\mathcal{D}q'$ as in \eqref{eq:fey_meas}, that is, with normalization $A_N$ and $A_N^*$.

The influence functional captures the dynamical effects of the traced-out environment on the system of interest. While exact solutions are typically limited to linear interactions~\cite{kamenev2023field}, approximate forms are necessary for general couplings. We consider an interaction of the form $H_{\text{int}} \propto f(q)Q$, where $f$ denote arbitrary function of the system and environment coordinates, respectively. By tracing out the environment degrees of freedom $Q$ and retaining terms up to second order in the coupling, we obtain the Gaussian influence functional \cite{calzetta1994noise}
\begin{widetext}
\begin{equation}
    \mathcal{F}[q,q'] = \exp \left( \frac{i}{2\hbar}\int dt \, dt' \Big[f(q)-f(q')\Big]_t D_{(t,t')} \Big[f(q)+f(q')\Big]_{t'} + \frac{1}{2}\int dt \, dt' \Big[f(q)-f(q')\Big]_t N_{(t,t')} \Big[f(q)-f(q')\Big]_{t'} \right).
    \label{eq:generalized_influence_functional}
\end{equation}
\end{widetext}
This expression generalizes the influence functional used in Ref.~\cite{calzetta1994noise}. Physically, it corresponds to a quadratic (Gaussian) approximation of the influence functional, valid for a non-linear coupling $f(q)Q$ to the environment. 
\clearpage
To bridge the quantum dynamics with classical physics, we introduce the mean coordinate $x$ (associated with the position of the particle in the classical limit) and the difference coordinate $y$ (associated with the quantum coherence length) \cite{weiss2012quantum}. We define
\begin{subequations}
\begin{align}
    &x = \frac{q+q'}{2}, \quad y = q-q', \\
    &q = x+ \frac{y}{2}, \quad q' = x - \frac{y}{2}.
\end{align}
\label{eq:coordinates_transformation}
\end{subequations}
The path integrals in $q,q'$ can be rewritten as path integrals in $x,y$, and such linear changes of variables have no Jacobian factors; we retain the $|A_N|$ normalization for $\mathcal{D}x$ and $\mathcal{D}y$. By assuming that the quantum fluctuations are small compared to the variation scale of the coupling, we can approximate the functional terms linearly in $y$
\begin{subequations}
\begin{align}
    f(q)-f(q') &= f\left( x+ \frac{y}{2}\right)-f\left(x - \frac{y}{2}\right) \approx f'(x) y, \\
    f(q)+f(q') &= f\left( x+ \frac{y}{2}\right)+f\left(x - \frac{y}{2}\right) \approx 2 f(x).
\end{align}
\label{eq:linear_expansion}
\end{subequations}
This linearization is physically justified by the scale separation. We assume that the characteristic scale $l$ of the variable $x$ is much larger than the coherence length $l_c$ (represented by $y$), such that $l_c/l \ll 1$. In this regime, the macroscopic nature of $x$ effectively suppresses quantum coherences. Although the influence functional is formally derived by integrating out the bath, strictly speaking, the validity of this strong decoherence limit depends on specific system parameters; here, however, we discuss it within a general framework.

By rewriting the influence functional, Eq.~\eqref{eq:generalized_influence_functional} in terms of $x_t$ and leading order in $y_t$, we have the influence functional in the strong decoherence approximation
\begin{equation}
    \mathcal{F}[x_{\tau},y_{\tau}] =\exp \left( \frac{i}{\hbar}\int dt \; dt' f'(x_t)f(x_{t'})y_t D_{(t,t')}  +\frac{1}{2}\int dt \; dt' y_t \left[f'(x_t)f'(x_{t'})N_{(t,t')}\right]y_{t'}  \right),\label{eq:gaussian influence}
\end{equation}
The exponent of the influence functional can be decomposed into a real part, associated with fluctuations (noise), and an imaginary part, associated with dissipation \cite{calzetta1994noise, paraguassu2024quantum}. The real part, responsible for decoherence, can be linearized by applying the Hubbard-Stratonovich transformation\footnote{This step, while not strictly mandatory, greatly simplifies the physical interpretation by introducing a stochastic variable.}. We introduce an auxiliary field $\eta_t$ via a Wiener path integral
\begin{equation}
    \mathcal{F}[x_{\tau},y_{\tau}] = \int \mathcal{D}\eta \exp\left(-\frac{1}{2}\int dt \, dt'\, \eta_t \left[ N_{(t,t')} \right]^{-1} \eta_{t'} \right) \exp \left( \frac{i}{\hbar}\int dt \left[ \int dt' f'(x_t) D_{(t,t')} f(x_{t'}) + \eta_t f'(x_t) \right] y_t \right),
    \label{eq:hubarstrat}
\end{equation}
where we have grouped the dissipative and stochastic terms acting on $y_t$. 

To ensure the rigorous normalization of the path integral, the functional measure $\mathcal{D}\eta$ must be carefully defined. For a Gaussian colored noise $\eta(t)$ characterized by a correlation kernel $N(t, t')$, the measure is constructed via time-slicing as
\begin{equation}
    \mathcal{D}\eta =\sqrt{\det\left(N^{-1}\right)}   \prod_{t=0}^\tau \frac{d\eta_t}{\sqrt{2\pi / dt}}.
    \label{eq:noise_measure}
\end{equation}
This distinction is crucial because, unlike the Feynman measure in Eq.~\eqref{eq:fey_meas}  the stochastic colored Wiener measure must explicitly account for the non-local temporal correlations through the functional determinant~\cite{chaichian2018path, hanggi1989path}. In the context of the influence functional, the prefactor $\sqrt{\det N^{-1}}$ acts as the necessary Jacobian for the Hubbard-Stratonovich transformation. This guarantees that the statistical weight of the noise trajectories is properly normalized to unity, 
\begin{equation}
    \int \mathcal{D}\eta \exp\left( -\frac{1}{2} \int_0^\tau \int_0^\tau \eta(t) N^{-1}(t,t') \eta(t') dt dt' \right) = 1,
\end{equation}
ensuring that the resulting dynamics for the reduced density matrix remains trace-preserving within the Quantum Induced Stochastic Dynamics (QISD) framework.

We now turn our attention to the system action and the superpropagator, performing a linearization with respect to the coherence coordinate $y_t$. Assuming the Lagrangian $L_q= \frac{1}{2}m \dot q^2 - V(q)$, we write
\begin{equation}
\int dt \left(L_q - L_q'\right) = \int dt \left[ m \dot x_t \dot y_t - \left(V(x_t+y_t/2) - V(x_t-y_t/2)\right) \right].
\end{equation}
Expanding to first order in $y_t$ and integrating by parts yields:
\begin{equation}
\int dt \left(L_q - L_q'\right) = (my_\tau v_\tau -my_0 v_0) -\int dt  y_t\left( m \ddot x_t + V'(x_t)\right),
\end{equation}
where we defined $\dot x(\tau) = v_\tau$ and $ \dot x(0) = v_0$. Note that, similar to the expansion of $f(q)$, the validity of this approximation depends on the specific form of the potential. For a quadratic (harmonic) potential, this expression is exact in $y_t$, whereas for general potentials, higher-order corrections depend on the degree of anharmonicity.

Let's go back to the superpropagator, bringing all ingredients together, now it is linear in $y_t$, and we have
\begin{eqnarray}
     \mathcal{J}[x_\tau,y_\tau|x_0, y_0] = \int dv_r dv_0 e^{\frac{i}{\hbar}m(y_\tau  v_r -y_0 v_0)}   \int_{v_0,v_\tau} \mathcal{D}\eta \; \mathcal{D}x \; \mathcal{D}y  \; \exp\left(-\frac{1}{2}\int dt \; dt'\; \eta_t \left[ N_{(t,t')}  \right]^{-1} \eta_{t'} \right) \\ \nonumber \times \exp \left( \frac{i}{\hbar}\int dt \left(- m \ddot x_t - V'(x_t)+ \int dt' f'(x_t)f(x_{t'})D_{(t,t')} + f'(x_t)\eta_t  \right)y_t \right),
\end{eqnarray}
where we integrate over all paths $x(t)$ with $x(0)=x_0$, $x(\tau)=x_\tau$, $\dot{x}(0)=v_0$ and $\dot{x}(\tau)=v_\tau$. The integral over the initial and final velocities $v_0,v_\tau$ means that we are integrating over all paths with $x(0)=x_0$ and $x(\tau)=x_\tau$.

The path integration over the coherence length $y_t$ results in a functional Dirac delta, enforcing the QISD equation of motion, this point is not new, as already showed in \cite{kamenev2023field, weiss2012quantum}. Explicitly, we have
\begin{align}
    \int \mathcal{D}y \, \exp &\left[ \frac{i}{\hbar}\int dt \, y_t \left(- m \ddot x_t - V'(x_t)+ \int dt' f'(x_t)D_{(t,t')} f(x_{t'}) + f'(x_t)\eta_t \right) \right] \nonumber \\
    &= \prod_{t=0}^\tau \left[ \frac{1}{\sqrt{\frac{2i\pi\hbar}{m}dt}} \left(\frac{2\pi\hbar }{ dt }\right) \right] \delta\left(m \ddot x_t + V'(x_t)- \int dt' f'(x_t)D_{(t,t')} f(x_{t'}) - f'(x_t)\eta_t\right).
\end{align}
The prefactors arise from the discretization of the path integral measure, ensuring that the constraint is satisfied at each time instant. Furthermore, to isolate the stochastic noise term $\eta_t$, we can rescale the argument of the Dirac delta using the property $\delta(aX) = \frac{1}{|a|}\delta(X)$. This yields
\begin{align}
    \prod_{t=0}^\tau &\left(\frac{2\pi\hbar }{ dt }\right)\delta\left(m \ddot x_t + V'(x_t)- \int dt' f'(x_t) D_{(t,t')} f(x_{t'})- f'(x_t)\eta_t\right) \nonumber \\
    &= \prod_{t=0}^\tau \left(\frac{2\pi\hbar }{ dt\, |f'(x_t)|}\right)\delta\left( \frac{1}{f'(x_t)}\left[ m \ddot x_t + V'(x_t)- \int dt' f'(x_t) D_{(t,t')} f(x_{t'}) \right] - \eta_t \right).
\end{align}
This transformation introduces the Jacobian determinant $|f'(x_t)|^{-1}$ into the measure. By evaluating the Wiener path integral over the noise variable $\eta_t$ via the Dirac delta functional, we obtain the explicit form of the superpropagator. To maintain a concise notation while preserving the measure factors, we define the dynamical functional $\mathcal{E}_t[x]$ as:
\begin{equation}
    \mathcal{E}_t[x] \equiv m \ddot x_t + V'(x_t)- \int ds \, f'(x_t)D_{(t,s)} f(x_s).
\end{equation}
Making the normalization factors of the path integrals explicitly, the superpropagator $\mathcal{J}$ can then be expressed as:
\begin{equation}
\begin{split}
    \mathcal{J}[x_\tau,y_\tau|x_0, y_0] &= \int dv_0 dv_\tau \, e^{\frac{i}{\hbar}m(y_\tau  v_\tau -y_0 v_0)} \int \mathcal{D}x \left[  \sqrt{\det\left(N^{-1}\right)} \prod_{t=0}^\tau  \frac{1}{\sqrt{2\pi /dt}}\frac{1}{\mathcal{N}_{\text{qm}}}\left(\frac{2\pi\hbar }{ dt |f'(x_t)|}\right) \right]   \\
    &\quad \times \exp \left[ -\frac{1}{2}\int dt \, dt' \frac{1}{f'(x_t)} \mathcal{E}_t[x] \, \left[ N_{(t,t')} \right]^{-1} \, \frac{1}{f'(x_{t'})} \mathcal{E}_{t'}[x] \right],
\end{split}
\end{equation}
where $\mathcal{N}_{\text{qm}} = \sqrt{\frac{2\pi i\hbar dt}{m}} \left(\sqrt{\frac{2\pi i\hbar dt}{m}}\right)^*$ represents the product of the forward and backward Feynman normalization constants. Let us explicitly analyze the combination of these measure factors
\begin{equation}
    \prod_{t=0}^\tau (dt)\,\underbrace{\frac{\sqrt{\det\left(N^{-1}\right)}}{\sqrt{2\pi dt}}}_{\text{Wiener}} \underbrace{\frac{dx_t}{\mathcal{N}_{\text{qm}}}}_{\text{Feynman}} \underbrace{\left(\frac{2\pi\hbar }{  dt |f'(x_t)|}\right)}_{\text{Jacobian}} 
    = \prod_{t=0}^\tau \left(\frac{m}{dt}\right)\frac{dx_t}{\sqrt{2\pi dt}|f'(x_t)|}\sqrt{\det\left(N^{-1}\right)}.
\end{equation}
At this point, it is crucial to interpret the physical significance of this result. We began with the Feynman-Vernon formalism, grounded in the oscillatory Feynman measure of quantum mechanics. However, through the strong decoherence limit the measure has fundamentally transformed. We have recovered a classical stochastic probability measure, weighted by the Jacobian $|f'(x_t)|^{-1}$ and the determinant of the kernel $N^{-1}(t,t')$. The same measure for a underdamped stochastic dynamics with multiplicative noise \cite{arenas2010functional} and colored noise. Absorbing the constants in $\mathcal{D}x$, we have the stochastic path integral is now

\begin{align}
    \mathcal{J}[x_\tau,y_\tau|x_0, y_0] &= \int dv_0 dv_\tau e^{\frac{i}{\hbar}m(y_\tau v_\tau -y_0 v_0)} \int \mathcal{D}x \;\exp \Bigg[ -\frac{1}{2}\int \frac{dt}{f'(x_t)} \, \frac{dt'}{f'(x_{t'})} \nonumber \\
    &~ \times \left( m \ddot x + V'(x)- \int ds f'(x_{t})f(x_{s})D_{(t,s)} \right)_t \left[ N_{(t,t')} \right]^{-1} \left( m \ddot x + V'(x)- \int ds f'(x_{t'})f(x_{s})D_{(t',s)} \right)_{t'} \Bigg].
\end{align}
The superpropagator takes the form of a Wiener path integral featuring a memory kernel $\left[ N_{(t,t')} \right]^{-1}$, a non-Markovian driving term $\sim D_{(t,s)}$, and boundary terms $y_\tau$ and $y_0$. To illustrate this formalism, we consider a concrete example. We note that the Hubbard-Stratonovich transformation is used here for simplicity, although equivalent results are obtainable by integrating the Gaussian influence functional of Eq.~\eqref{eq:gaussian influence} directly.

\subsection{Example: High-Temperature Caldeira-Leggett Model}
The Caldeira-Leggett model considers a linear interaction between a system and an ensemble of harmonic oscillators. By tracing out the oscillators in a thermal state and taking the high-temperature limit, the model recovers classical Brownian motion~\cite{caldeira1983path} and is exact. Therefore, it serves as an ideal benchmark to bridge the gap between quantum and classical path integrals. In this limit, the dissipation and noise kernels, along with the linear coupling function, are given by
\begin{equation}
    D_{(t,s)} = \gamma \frac{d}{dt}\delta(t-s), \quad N_{(t,t')} = 2\gamma k_B T \delta(t-t'), \quad f(x) = x. 
\end{equation}
Note that for linear coupling $f(x)=x$, the Jacobian term derived in the previous section becomes unity ($|f'(x)|=1$), while the determinant of the kernel becomes $\sqrt{\det\delta}\rightarrow 1$, simplifying the measure. Substituting these kernels into our general expression for the superpropagator, we obtain:
\begin{align}
    \mathcal{J}[x_\tau,y_\tau|x_0, y_0] &= \int dv_\tau \int dv_0\;e^{\frac{i}{\hbar}m(y_\tau v_\tau -y_0 v_0)} \int \mathcal{D}x \;\exp \Bigg[ -\frac{1}{4\gamma k_B T}\int {dt}  \left( m \ddot x + V'(x)+\gamma \dot x  \right)^2  \Bigg].
\end{align}
Inspecting the exponent, we recognize exactly the Onsager-Machlup action for an underdamped Brownian motion~\cite{paraguassu2023heat}. This result exemplifies how the Feynman-Vernon formalism, rooted in quantum mechanics, naturally converges to the Wiener path integral formalism of stochastic processes.

\section{Stochastic evolution of the Wigner Function}\label{sec: estochastic evol}
The Wigner function serves as the natural bridge between quantum and classical dynamics. By employing the phase-space formulation of quantum mechanics, one can rigorously trace the emergence of classicality~\cite{zachos2005quantum}. An immediate consequence of the super-propagator derived in the previous section is that the time evolution of the Wigner function is governed by a stochastic path integral, a connection we explicitly demonstrate in this section.

We start with the density matrix in terms of $x$ and $y$ coordinates
\begin{equation}
    \rho_\tau\left(x_\tau+\frac{y_\tau}{2},x_\tau-\frac{y_\tau}{2}\right)  = \int dx_0 \int dy_0 \;\rho_0\left(x_0+\frac{y_0}{2},x_0-\frac{y_0}{2}\right) \mathcal{J}[x_\tau,y_\tau|x_0, y_0],
\end{equation}
We go to the Wigner description by taking the Fourier transform in $y_\tau$. Going from $y_\tau$ to $p_\tau$, we have
\begin{equation}
    W(x_\tau, p_\tau) = \int dy_\tau \rho\left(x_t+\frac{y_t}{2} , x_t - \frac{y_t}{2} \right) e^{- \frac{i}{\hbar}y_t p_t} = \int dy_\tau \; e^{- \frac{i}{\hbar}y_t p_t}\int dx_0 \int dy_0 \;\rho\left(x_0+\frac{y_0}{2},x_0-\frac{y_0}{2}\right) \mathcal{J}[x_\tau,y_\tau|x_0, y_0] 
\end{equation}
The initial density matrix can also be written in terms of the Wigner function, by inverse Fourier transform
\begin{eqnarray}
    \rho\left(x_0+\frac{y_0}{2} , x_0 - \frac{y_0}{2} \right) = \int \; dp_0 \;e^{\frac{i}{\hbar}y_0 p_0}\; W(x_0, p_0)
\end{eqnarray}
so we have
\begin{eqnarray}
     W(x_\tau, p_\tau) = \int dy_\tau \; e^{- \frac{i}{\hbar}y_t p_t}\int dx_0 \int dy_0 \;\int \; dp_0 \;e^{\frac{i}{\hbar}y_0 p_0}\; W(x_0, p_0) \mathcal{J}[x_\tau,y_\tau|x_0, y_0]\label{eq:wignerfunction1}
\end{eqnarray}
Moreover the superpropagator can be written as
\begin{equation}
    \mathcal{J}[x_\tau,y_\tau|x_0, y_0] = \int dv_0 \int dv_\tau \; e^{\frac{i}{\hbar}m(y_\tau  v_\tau -y_0 v_0)} P[v_\tau, x_\tau| v_0, x_0], 
\end{equation}
where $P[v_\tau, x_\tau| v_0, x_0]$ is the stochastic conditional distribution probability for an underdamped stochastic process \cite{paraguassu2023heat}
\begin{equation}
   P[v_\tau,x_\tau|v_0,x_0]= \int \mathcal{D}x \;\exp \Bigg[ -S_{s}[x] \Bigg],
\end{equation}
\begin{eqnarray}
    S_{s}[x] &=& \frac{1}{2}\int \frac{dt}{f'(x_t)} \, \frac{dt'}{f'(x_{t'})} \\
    && \quad \times \left( m \ddot x + V'(x)- \int ds f'(x_{t})f(x_{s})D_{(t,s)} \right)_t \left[ N_{(t,t')} \right]^{-1} \left( m \ddot x + V'(x)- \int ds f'(x_{t'})f(x_{s})D_{(t',s)} \right)_{t'} \nonumber
\end{eqnarray}
Bringing all ingredients together, the Wigner function, Eq.~\eqref{eq:wignerfunction1}, becomes
\begin{equation}
    W(x_\tau,p_\tau) = \int dy_\tau \; e^{- \frac{i}{\hbar}y_t p_t}\int dx_0 \int dy_0 \;\int \; dp_0 \;e^{\frac{i}{\hbar}y_0 p_0}\; W(x_0, p_0) \int dv_0 \int dv_\tau \; e^{\frac{i}{\hbar}m(y_\tau  v_\tau -y_0 v_0)} P[v_\tau, x_\tau| v_0, x_0]. 
\end{equation}
The integrals in $y'$s give a Dirac delta function, and we find
\begin{equation}
    W(x_\tau,p_\tau) =\int dv_\tau \int dv_0 \int dp_0 \;\delta(m v_\tau - p_\tau) \delta(m v_0 - p_0) W(x_0, p_0) P[v_\tau, x_\tau| v_0, x_0] 
\end{equation}
The effect of the Dirac delta is to force $p=mv$, by integrating out the velocities, we finally have
\begin{equation}
     W(x_\tau,p_\tau) = \int dx_0 dp_0  W(x_0, p_0) P[p_\tau, x_\tau| p_0, x_0] .\label{eq:wigner function}
\end{equation}
This result is of central importance: it demonstrates that in the strong decoherence limit, we arrive at a simple composition of probabilities. In this regime, the dynamics of $W$ is indistinguishable from the evolution of a classical probability satisfying a standard stochastic process. Thus, $W$ represents a proper probability distribution whenever the initial Wigner function is well-defined as such. In particular, this framework allows one to investigate the decoherence of nonclassical initial states—such as Fock states—through a stochastic path integral formulation.

\section{Quantization of a Stochastic Processes}\label{sec: quantization}

The correspondence established between the Feynman and Wiener measures in Sec.~\ref{sec: feynmantowiener} constitutes a bidirectional bridge between the quantum and classical domains. Just as this mapping allowed us to derive the stochastic evolution of the Wigner function in the previous section, it naturally enables the investigation of the inverse problem: the quantization of a classical stochastic process.

By reversing the logic applied so far, we can construct the quantum influence functional associated with a known classical stochastic differential equation. This phenomenological approach yields a valid dynamical description of an open quantum system without requiring knowledge of the microscopic details of the environment. This framework is particularly powerful in scenarios where classical stochastic phenomena are well-characterized, allowing one to explore their corresponding quantum regimes by postulating the structure of the influence functional.

Consider a generalized stochastic differential equation with multiplicative noise
\begin{equation}
    m \ddot{x} + V'(x) + \int_{0}^{t} dt' \, g(x_{t'}) D(t,t') = f(x_t)\xi(t),
    \label{eq:langevin}
\end{equation}
where $\xi(t)$ is a Gaussian noise term with zero mean and correlation function $\langle \xi(t) \xi(t') \rangle = N(t,t')$. The functions $f(x)$ and $g(x)$ characterize the state-dependent fluctuations and dissipation, respectively.

To quantize this system, we construct the stochastic generating functional (the classical path integral) by enforcing the equation of motion via a Dirac delta constraint. Using the integral representation of the delta function with an auxiliary variable $y_t$ (which maps to the quantum difference coordinate), the functional reads
\begin{align}
    \mathcal{F}[x,y] &= \int \mathcal{D}\xi \; e^{-\frac{1}{2}\int dt \int dt' \xi(t) N^{-1}(t,t') \xi(t')} \; \exp\left[ \frac{i}{\hbar} \int dt \, y_t \Big( f(x_t)\xi(t) - \int dt' g(x_{t'}) D(t,t') \Big) \right] .
    \label{eq:influence_func_raw}
\end{align}
Since $\xi(t)$ is Gaussian, the average can be computed exactly. Integrating out the noise yields the effective influence functional:
\begin{equation}
    \mathcal{F}[x,y] = \exp\left( -\frac{1}{2\hbar^2} \int dt \, dt' \, y_t f(x_t) N(t,t') f(x_{t'}) y_{t'} - \frac{i}{\hbar} \int dt \, dt' \, y_t g(x_t)D(t,t') \right).
    \label{eq:influence_func_final}
\end{equation}
This expression is precisely the generalized Feynman-Vernon influence functional derived in Eq.~\eqref{eq:generalized_influence_functional}. Thus, any classical stochastic process described by Eq.~\eqref{eq:langevin} has a direct open quantum system analog. Moreover, following the methods of Ref.~\cite{PhysRevD.45.2843}, one can use this functional to derive a non-Markovian master equation governing the system's dynamics.

It is crucial to distinguish the present framework from stochastic formulations such as Parisi-Wu Stochastic Quantization \cite{damgaard1987stochastic} and Nelson's Stochastic Mechanics \cite{nelson1966derivation}. Unlike these approaches, which either introduce fictitious time or postulate universal stochasticity to derive quantum mechanics, our method remains rooted in the real-time dynamics of open quantum systems. Here, stochasticity is neither an axiom nor a computational device, but an emergent property of the Feynman-Vernon influence functional. Consequently, our 'inverse problem' is phenomenological: we identify noise and dissipation kernels that correspond to a specific classical Langevin equation. Once reconstructed, the influence functional enables a complete quantum analysis beyond the semiclassical approximation, capturing non-classical features and coherence even in non-harmonic potentials.

\section{Conclusion}\label{sec: conclusion}
In conclusion, we have established a  mathematical link between the Feynman-Vernon formalism of open quantum systems and the Wiener path integral of classical stochastic theory. By enforcing the strong decoherence limit, we demonstrated that the oscillatory Feynman measure naturally transforms into the probabilistic Wiener measure. In doing so, we have formalized the framework of Quantum Induced Stochastic Dynamics (QISD), providing a rigorous footing for treating quantum fluctuations as effective classical stochastic sources.

This correspondence has two immediate consequences. On the one hand, it demonstrates that the time evolution of the Wigner function can be reconstructed from classical stochastic trajectories—which are exact for linear system–environment couplings—ensuring that the Gaussian effective description recovers standard probability properties. On the other hand, it provides a concrete formulation of the inverse problem, yielding a systematic procedure to quantize classical stochastic dynamics. This creates a robust pathway to investigate decoherence mechanisms in non-Markovian baths.

Beyond the formal aspects discussed here, this framework naturally raises questions about the energetic interpretation of the resulting quantum-induced trajectories. In particular, it offers a suitable setting to investigate stochastic energetics and thermodynamic fluctuations at the quantum–classical interface. Moreover, the inverse quantization scheme can be generalized to construct effective Langevin descriptions for more complex systems, with promising applications in regimes characterized by multiplicative noise.

\section{Acknowledges}

We acknowledge Luca Abrahão, Thiago Guerreiro and  Nami Svaiter for useful discussions. All authors acknowledge the Funda\c{c}\~ao de Amparo \`a Pesquisa do Estado do Rio de Janeiro, FAPERJ Process SEI-260003/000174/2024 (P.V.P.) and PhDs merit fellowship - FAPERJ Nota 10, 203.709/2025 (F.S.) and 201.622/2025 (B.S.). This work was partially supported by StoneLab.

\providecommand{\newblock}{}
\bibliography{name}

\end{document}